# Dissociation and ionization equilibria of deuterium fluid over a wide range of temperatures and densities

Mofreh R. Zaghloul

Department of Physics, College of Science, United Arab Emirates University, P.O. Box 15551, Al-Ain, UAE.

## ABSTRACT

We investigate the dissociation and ionization equilibria of deuterium fluid over a wide range of temperatures and densities. The partition functions for molecular and atomic species are evaluated, in a statistical-mechanically consistent way, implementing recent developments in the literature and taking high-density effects into account. A new chemical model (free energy function) is introduced in which the fluid is considered as a mixture of diatomic molecules, atoms, ions and free electrons. Intensive short range hard core repulsion is taken into account together with partial degeneracy of free electrons and Coulomb interactions among charged particles. Samples of computational results are presented as a set of isotherms for the degree of ionization, dissociated fraction of molecules, pressure, and specific internal energy for a wide range of densities and temperatures. Predictions from the present model calculations show an improved and sensible physical behavior compared to other results in the literature.

### I – INTRODUCTION

The equation-of-state of deuterium, over a wide range of densities and temperatures, is of central importance to many fundamental and applied topics and has been under intensive investigation over the past four decades [1-31]. Accurate prediction of the equation-of-state and thermodynamic properties of deuterium at ultrahigh pressures and a wide range of temperatures is of great interest to ICF research and astrophysical applications. Since the compressibility of a material is determined by its equation-of-state, the accurate estimation of the EOS of the deuterium or deuterium-tritium fusion fuel is critical for designing ICF ignition targets and for predicting the performance of the target during ICF implosions. Besides, both of the planets Jupiter and Saturn contain large amount of very dense hydrogen and information about deuterium EOS is, therefore, essential in the study of these planets.

Results of shock-wave experiments on fluid hydrogen and other diatomics in the 100 GPa range have shown metallization (insulator to conductor transition) at densities as high as tenfold of initial liquid density, and temperatures from a few to several thousand K [32-36]. Precise estimation of the properties of such diatomics considering dissociation and ionization equilibria with special interest given to ultrahigh pressure phenomena such as pressure dissociation and pressure ionization is crucial for the understanding and explanation of these experimental results.

In the present paper, and within the chemical picture, we present a free energy model for the calculation of the EOS and thermodynamic properties of deuterium over a wide range of densities and temperatures. Dissociation and ionization equilibria are considered with a special interest devoted to the pressure dissociation and pressure ionization phenomena. The model implements our newly proposed formulation for the establishment of statistical-thermodynamically-consistent finite bound-state partition functions [37-39] where an occupational probability is prescribed in advance to assure a smooth *truncation* of the electronic partition function and manifestation of the phenomenon of pressure ionization. An analogous occupational probability is proposed herein for a smooth truncation of the vibrational-rotational molecular partition function of deuterium molecules. Quantum effects, in electrons free energy, together with Coulomb and excluded-volume configurational terms are taken into account and the resulting free energy function is then used to calculate the occupation numbers and to find the corresponding set of nonideal thermodynamic properties.

It is worth mentioning that in the calculation of the internal electronic partition function using occupational probabilities, one has to avoid cutting-off the ground states which causes the nonphysical vanishing of the partition function; a problem that has been recently scrutinized showing the failure of the so-called Planck-Larkin partition function at high temperatures and the failure of the state-dependent partition functions with *commonly used* occupational probabilities as they vanish at low temperatures and high densities [40]. A remedy proposed in Ref. [40] is used herein to avoid this problem and is applied to the conventional rotational-vibrational molecular partition functions as well.

At equilibrium, the free energy function must be a minimum and the equilibrium composition can be obtained, therefore, through the solution of the resulting set of minimization equations subject to conservation of electric charge and the stoichiometric constraints. This can be done by using numerical optimization algorithms [41-44] or through casting the resulting minimization equations analytically into the form of nonideal (or modified) Saha equations and solving the resulting set of nonlinear algebraic equations subjected to the above mentioned constraints using efficient algorithms developed in the literature [45-47]. Once the equilibrium composition is determined one can calculate the thermodynamic functions from the equilibrium free energy using the standard thermodynamic relations.

## II– DISSOCIATION AND IONIZATION EQUILIBRIA

At high temperatures fluid deuterium is, in general, a complex mixture of diatomic molecules ($D_2$), neutral atoms ($D$), in addition to deuterons ($D_+$) and free electrons ($e$). For such an assembly, statistical theory, with some assumptions, makes it possible to express the bulk-state functions in terms of the relevant properties of the individual particles. Interactions among different species are commonly taken into account through considering a configurational free energy term and a scheme for truncating the internal partition function. It is unlikely, however, that molecular ions exist in high concentrations that affect the equation-of-state and they can be readily neglected. Ignoring the formation of molecular ions simplifies the task as thus a molecule cannot ionize without being dissociated. For such a mixture of molecular deuterium, $D_2$, atomic deuterium, $D$, deuterons, $D_+$, and free electrons, $e$; the free energy function of the system can be approximated as the sum of separable components;

$$F = F_{D_2} + F_D + F_{D+} + F_e + F_{corr} + F_{zero-of-energy}, \quad (1)$$

where $F_{corr}$ is a term that takes into account possible configurational and quantum mechanical (degeneracy) effects and the "*zero-of-energy*" term, $F_{zero\text{-}of\text{-}energy}$, takes into count and corrects for the fact that the free energy components in Eq. (1) must all be calculated using energies referred to the same reference or "*zero-of-energy*". In accordance with the usual practice, the *zero-of-energy* to which all other energies are referred to will be taken as the energy of dissociated atoms at infinite distance apart in their ground states.

At very high temperatures, the contribution of the photon gas to the thermodynamics of the plasma becomes crucial and one has to add to the right hand side of Eq. (1) the free energy of the photon gas which can be expressed, for a blackbody radiation, as

$$F_{bb_rad} = -(4\sigma/3c)\, V\, T^4 \quad (2)$$

where $\sigma$ is the Stefan-Boltzmann constant, $T$ is the absolute temperature, and $c$ is the speed of light.

Among different kinds of particles there may exit a set of chemical reactions including dissociation, ionization, and their inverse recombination processes. At equilibrium, the free energy function (Eq. (1)) is minimized for each of the possible reactions. Minimizing the free energy for the dissociation process $D_2 \leftrightarrow 2D$ requires that

$$\frac{\partial F}{\partial N_{D_2}} - 2\frac{\partial F}{\partial N_D} = 0 \quad (3)$$

While the minimization of *F* for the ionization equilibrium reaction $D \leftrightarrow D_+ + e$ requires that

$$\frac{\partial F}{\partial N_D} - \frac{\partial F}{\partial N_{D_+}} - \frac{\partial F}{\partial N_e} = 0 \qquad (4)$$

In Eqs. (3,4) $N_{D_2}$, $N_D$, $N_{D+}$, and $N_e$ stand for the occupation numbers of deuterium molecules, atomic deuterium, deuterons and free electrons, respectively.

Solving Eqs (3) and (4) subject to conservation of electric charge and the stoichiometric constraints gives the required equilibrium composition. Once the equilibrium composition is determined, one can proceed to calculate the set of thermodynamic properties using standard thermodynamic relations. However, explicit expressions for the components of the free energy in Eq. (1) are needed for the accomplishment of this task. In the following section (section III) we present the set of expressions for the *classical* ideal components of the free energy while the configurational and quantum mechanical effects are discussed in section IV.

## III– CLASSICAL-IDEAL FREE ENERGY COMPONENTS

The classical-ideal free energy component for molecular deuterium can be written as

$$F_{D_2} = -K_B T N_{D_2} \left( \ln\left( \frac{Q_{tot,D_2}}{N_{D_2}} \right) + 1 \right) \qquad (5)$$

where the total partition function for the molecules, $Q_{tot,D_2}$, can be factored as

$$Q_{tot,D_2} = \left( \frac{2\pi m_{D_2} K_B T}{h^2} \right)^{3/2} V\, Q_{vib}\, Q_{rot}\, Q_{elec,D_2} \qquad (6)$$

The first factor in the right hand side of the above equation represents the translational partition function while $Q_{vib}$, $Q_{rot}$ and $Q_{elec,D_2}$ are the vibrational, rotational and electronic factors, respectively. Because diatomic molecules can exist in appreciable amounts only if the temperature is small compared with the dissociation energy of the molecule, retaining the ground electronic state of the molecule only may be regarded as a reasonable approximation. For *homonuclear* diatomic molecules like $D_2$, the rotational levels are divided into *ortho* and *para* series characterized, in order, by odd and even numbered rotational levels and the vibrational-rotational factor, $Q_{vib}Q_{rot}$, can be written as [48]

$$Q_{vib}Q_{rot} = \sum_{n}\sum_{\text{even}\,J}(2J+1)\exp\left(-\frac{[E_n+E_J]}{K_BT}\right)+3\sum_{n}\sum_{\text{odd}\,J}(2J+1)\exp\left(-\frac{[E_n+E_J]}{K_BT}\right) \tag{7}$$

where $n=0,1,2,\ldots$ is the vibrational quantum number, $J=0,1,2,\ldots$ is the rotational quantum number, $E_n=-D_e+(n+½)hv$ is the $n^{\text{th}}$ vibrational energy of a harmonic oscillator and $E_J=-D_e+J(J+1)h^2/8I\pi^2$ is the $J^{\text{th}}$ rotational energy of a rigid rotor where $I$ is the moment of inertial of the molecule, $h$ is the Planck's constant and $v$ is the vibrational frequency. The energies $E_n$ and $E_J$ are referred to the above mentioned common *zero-of-energy* and $-D_e$ is the energy of the bottom of the well relative to the common *zero-of-energy*. It has to be noted that the rotational energy $E_J$ of the quantum number $J$ generally depends on $n$ when vibrational-rotational coupling is taken into account. Extending the summation in Eq. (7) from $n=0$ to $\infty$ and from $J=0$ to $\infty$ would imply the possibility of getting infinite rotational and vibrational energies. However, the atoms forming the molecule would dissociate before these extremes are reached. Accordingly, a cut-off scheme to truncate the summation in Eq. (7) is needed. A good approximation of the cut-off limit is the lowered dissociation energy. We incorporate this cut-off procedure in a smooth way using a smooth occupational probability function, $w_{n,J}(V,T,\{N\})$, similar to the one being used with the electronic partition function of atoms. Details about $w_{n,J}(V,T,\{N\})$are given in Section V. Taking the anharmonicity, centrifugal distortion, and vibrational-rotational coupling into account, the total rotational-vibrational energy levels, measured from the lowest, may be approximated by [26,48,49]

$$\varepsilon_{nJ} \approx \tilde{\nu}_e n + \tilde{B}_e J(J+1) - \tilde{\chi}_e \tilde{\nu}_e (n^2+n) - \tilde{D}_e J^2 (J+1)^2 - \alpha_e (n+\tfrac{1}{2}) J(J+1) \tag{8}$$

Values of the parameters $\tilde{\nu}_e, \tilde{B}_e, \tilde{\chi}_e \tilde{\nu}_e,\ \tilde{D}_e$ and $\alpha_e$ along with the dissociation energy, $D_0$, for deuterium are taken from Huber and Herzberg [49] and are given in Table 1.

Table 1. Spectroscopic parameters for the deuterium molecule.

| Molecule | $\tilde{\nu}_e$ [cm$^{-1}$] | $\tilde{\chi}_e \tilde{\nu}_e$ [cm$^{-1}$] | $\tilde{B}_e$ [cm$^{-1}$] | $\tilde{D}_e$ [cm$^{-1}$] | $\alpha_e$ [cm$^{-1}$] | $D_0$ [eV] |
|---|---|---|---|---|---|---|
| $D_2$ | 3115.5 | 61.82 | 30.44 | $1.141\times10^{-2}$ | 1.0786 | 4.55632 |

Considering the vibrational-rotational occupation probability and the remedy proposed in Ref. [40] together with the anharmonicity, centrifugal distortion, and vibrational-rotational coupling, one can write the vibrational-rotational factor as

$$\begin{aligned} Q_{vib}Q_{rot} &= e^{\frac{D_0}{K_B T}} \bar{Q}_{vib}\bar{Q}_{rot} \\ &= e^{\frac{D_0}{K_B T}} \Bigg[ [1 - w_{0,0}] + \sum_n \sum_{\text{even}\, J} w_{n,J}(V,T,\{N\})(2J+1)\exp\left(-\frac{\varepsilon_{nJ}}{K_B T}\right) \\ &\quad + 3\sum_n \sum_{\text{odd}\, J} w_{n,J}(V,T,\{N\})(2J+1)\exp\left(-\frac{\varepsilon_{nJ}}{K_B T}\right) \Bigg] \end{aligned} \tag{9}$$

where $D_0 = (D_e - \tilde{\nu}_e/2 + \tilde{\chi}_e \tilde{\nu}_e/4)$ is the dissociation energy of the molecule and the energies $\varepsilon_{nJ}$ are relative to the corresponding ground state (lowest). To ensure the correct behavior of the energy levels, the sum over $J$ is restricted to values for which $\left(\partial\varepsilon_{nJ}/\partial J\right)_n \geq 0$ [26]. In Eq. (9) $\overline{Q}_{vib}\overline{Q}_{rot}$ which is equal to the quantity in the square brackets represents the *conventional* (calculated from the ground state) vibrational-rotational factor of the partition function while the leading exponential can be considered as the corresponding *zero-of-energy* correction. This *zero-of-energy* factor is in every respect equivalent to a contribution to the free energy function given by

$$F_{D_2, zeo-of-energy} = -N_{D_2} D_0 \tag{10}$$

The classical-ideal free energy of atomic deuterium is given by a classical expression similar to that of $D_2$ where

$$F_D = -K_B T N_D \left( \ln\left( \frac{Q_{tot,D}}{N_D} \right) + 1 \right) \tag{11}$$

where the total partition function for deuterium atoms, $Q_{tot,D}$, is written as

$$Q_{tot,D} = \left( \frac{2\pi m_D K_B T}{h^2} \right)^{3/2} V \cdot Q_{elec,D} \tag{12}$$

The first factor in the right hand side of Eq. (12) is the translational partition function while $Q_{elec,D}$ is the electronic partition function of atomic deuterium. The evaluation of the atomic electronic partition function has been thoroughly studied and scrutinized in Refs [37-40] with particular interest devoted to the case of hydrogen. To avoid the cut-off of all ground states which causes the nonphysical vanishing of the electronic partition function we use the expression proposed in [40] where

$$Q_{elec}(V,T,\{N\}) = \left[1 - w_0(V,T,\{N\})\right] + \sum_{i=0}^{\infty} g_i w_i(V,T,\{N\}) e^{-\varepsilon_i/K_B T} \tag{13}$$

where $Q_{elec}(V,T,\{N\})$ is the conventional electronic partition function with energy states relative to the electronic ground state (which coincides here with the common *zero-of-energy*) and $g_i$ is the statistical weight of $i^{th}$ level whose energy relative to the ground is $\varepsilon_i$. The state-dependent occupational probability, $w_i(V,T,\{N\})$, of the $i^{th}$ level is taken from Salzmann, Ref. [50] where it decreases continuously and monotonically as the density (or the strength of relevant interaction) increases in order to produce a physically reasonable continuous transition between bound and free states. It also provides a natural and smooth truncation of the internal electronic partition function as it falls strongly to zero as the binding energy of a level below the unperturbed continuum goes to zero. It is very clear from Eq. (13) that the remedy proposed in Ref. [40] retains at least one state of the ground level in a strongly perturbed system where $w_0(V,T,\{N\})$ goes to zero at low temperatures.

Deuterons, on the other hand, have no internal electronic structure and, thus, their classical-ideal free energy can be written as

$$F_{D_+} = -K_B T N_{D+} \left[ \ln\left( \frac{V}{N_{D_+}} \left( \frac{2\pi m_{D_+} K_B T}{h^2} \right)^{3/2} \right) + 1 \right] \tag{14}$$

While the classical-ideal free energy for free electrons is expressed as

$$F^{cls}_{e,id} = -K_B T N_e \left[ \ln\left( \frac{2V}{N_e} \left( \frac{2\pi m_e K_B T}{h^2} \right)^{3/2} \right) + 1 \right] \tag{15}$$

Recalling that all interconnected energy levels must essentially be referred to the same reference, hence to the sum of $F_{D+}$ and $F_e$ one has to add the term

$$F_{e-D+,zero-of-energy} = N_{D_+}\ \chi_D \tag{16}$$

to account for our chosen *zero-of-energy*, where $\chi_D$ is the ionization energy of neutral deuterium atom.

## IV– DEGENERACY AND CONFIGURATIONAL CORRECTIONS

Quantum mechanical effects (partial degeneracy) and configurational corrections are briefed in Eq. (1) in the term, $F_{corr}$. This term can be expanded into the contributions

$$F_{corr} = F_{hs} + F_C + F_{bn} + F_{dgc,\eta} \tag{17}$$

where $F_{hs}$ is the free energy corrections due to the consideration of particles of finite volumes as a mixture of hard spheres of different diameters, $F_C$ is the Coulombic excess free energy which takes into account the interactions among charged particles, $F_{bn}$ is the correction due to the exclusion of the occupied volume from that accessible to bare nuclei (of vanishing diameters), and $F_{dgc,\eta}$ is the quantum mechanical (degeneracy) correction, with excluded volume taken into account, for free electrons, $F_{dgc,\eta}$.

For a mixture of diatomic molecules and neutral atoms, the hard core free energy can be written as [51,52]

$$F_{hs} = K_B T((1-f)N_0 + 2fN_0\alpha_0)\left[\frac{X\eta}{(1-\eta)^2} + \frac{3Y\eta}{1-\eta} + (X-1)\ln(1-\eta)\right] \tag{18}$$

where $N_0$ is the original number of molecules in the absence of dissociation, $f$ is the fraction dissociated from the original molecules while $\alpha_0$ is the proportion of neutral atoms to the total number of non-molecular heavy particles (sum of numbers of deuterons and neutral atoms). The packing parameter $\eta$ is given by

$$\eta = \frac{\pi}{6V}\left((1-f)N_0 d_m^3 + 2fN_0\alpha_0 d_0^3\right) = \frac{N_0\pi}{3V}\left((1-f)+f\alpha_0\right)d_0^3 \qquad (19)$$

where $d_m$ is the diameter of deuterium molecule. In the above expression for $\eta$, it is assumed that the volume of the molecule is the sum of the volumes of its constituent atoms. The parameters $X$ and $Y$ in Eq. (18) are given by

$$X = \left(\overline{d^2}\right)^3 \Big/ \left(\overline{d^3}\right)^2 \qquad (20)$$

and

$$Y = \left(\overline{d^2}\right)\left(\overline{d^1}\right) \Big/ \left(\overline{d^3}\right) \qquad (21)$$

with

$$\overline{d^k} = \tfrac{1}{2}(1-f)d_m^k + f\alpha_0 d_0^k \qquad (22)$$

The second term in the right hand side of Eq. (17) refers to the Coulomb correction to the free energy function. It is commonly known that exact methods for solving the many-particle problem of Coulomb-interacting systems are still missing [51,53,54] and, as a result, only approximate nonideal plasma models are routinely used. Here we adopt a cellular model in which the system is regarded as a sum of electrically neutral spherical unit cells where ions located at the centers of the cells simultaneously neutralize the background electrons. The Coulomb excess free energy derived in this case can be written as

$$F_C = -\frac{18}{10}\frac{N_0 f(1-\alpha_0)e^2}{4\pi\varepsilon_0 R_0} \qquad (23)$$

where the cell radius $R_0$ is given by $R_0 = (4\pi(1-\eta_m)V/6fN_0)^{1/3}$ with $\eta_m$ is the fraction of the volume occupied by non-dissociated molecules and $N_0$ is the original number of molecules.

The third term in the right hand side of Eq. (17) refers to the correction to the free energy function of bare nuclei or deuterons due to reduction in the accessible volume. Bare nuclei or deuterons can be regarded as point-like particles. The existence of extended particles, like atoms and molecules in our case, reduces the volume accessible to bare nuclei (deuterons). The exclusion of the occupied volume from the volume accessible to bare nuclei gives rise to a correction to the free energy that can be expressed as

$$F_{bn} = -K_B T N_{D+} \ln(1-\eta) \tag{24}$$

The correction given by Eq. (24) is equivalent to replacing the volume $V$ in Eq. (14) by a reduced volume $V^* = V(1-\eta)$ where $\eta$ is the packing parameter given by Eq. (19).

The last term in the right hand side of Eq. (17) refers to the corrections to the free energy function due to quantum effects and reduction in the accessible volume for free electrons. For Boltzmann non-degenerate particles one may use a correction term, similar to Eq. (24), to account for the reduction in volume in the ideal contribution of electrons. However, at high densities, electrons become partially or totally degenerate and quantum effects must be taken into account. Following Ebeling et al [51] and Kahlbaum and Forster [52], the free energy for free electrons with quantum effects and excluding the volume occupied by extended particles can be written as

$$F_{e,id}^{dgc} = -\left(2 K_B T V^* / \Lambda_e^3\right) I_{3/2}(\mu_{e,id}^* / K_B T) + N_e \mu_{e,id}^* \tag{25}$$

where $\mu_{e,id}^* = \partial F_{e,id}^{dgc} / \partial N_e$ is the chemical potential of the ideal Fermi electron gas, $V^* = V(1-\eta)$ is the reduced volume, and $I_\nu$ is the complete Fermi-Dirac integral. The electron chemical potential

$\mu^{*}_{e,id}$ is related to the number of free electrons in the system by

$$N_e = \left(2V^{*}/\Lambda_e^3\right) I_{1/2}(\mu^{*}_{e,id}/K_B T) \qquad (26)$$

where $\Lambda_e = \dfrac{h}{\sqrt{2\pi m_e K_B T}}$ is the average thermal wave length of the electrons. The densities per volume e.g. $f_{e,id}=F_{e,id}/V$ and $n_e=N_e/V$ are taken always according to the total volume $V$.

Replacing the classical expression, $F^{cls}_{e,id}$, Eq. (15), by the quantum expression, Eq. (25), the final expression for the free energy function will be the same as that for classical plasma plus a correction term, $F_{dgc,\eta} = F^{dgc}_{e,id} - F^{cls}_{e,id}$, to be considered with the other above mentioned free energy corrections.

## V– OCCUPATIONAL PROBABILITIES AND TRUNCATION OF PARTITION FUNCTIONS

The main message of Refs [37,38] was the introduction of a mathematically consistent formulation for calculating thermodynamic properties of plasma systems and establishing statistical-thermodynamically-consistent finite bound state partition function (BSPF) within the chemical picture. As has been shown in these references, this can be achieved through solving the problem in which, based on physical arguments, occupational probabilities are prescribed in advance to assure smooth truncation of the electronic partition functions and to demonstrate the phenomenon of pressure ionization. As direct consequences of using such a formulation and the scheme used, in advance, to truncate the BSPF, new correction terms will emerge in the expressions routinely used for calculating plasma thermodynamic functions as explained in Refs [37,38].

Many models for the occupational probability exist in the literature ranging in their level of sophistication. In the present calculation we use a simple model for the occupational probability given by Salzmann in Ref. [50] in which the occupational probability for energy level $i$ is given by

$$w_i = \exp\left[-\left(\frac{\Delta\chi_D}{\chi_{D,i}}\right)^3\right] \tag{27}$$

where $\chi_{D,i}$ is the energy of level $i$ measured from the continuum and $\Delta\chi_D$ is the shifting in the ionization potential of the neutral deuterium (negative energy) due to the configurational component of the free energy. It is evident from the above expression that $w_i$ goes to unity as $\Delta\chi_D$ goes to zero which is the case of ideal plasma and that $w_i$ smoothly goes to zero as $\Delta\chi_D$ exceeds $\chi_{D,i}$ (in magnitude) as desired for the truncation of the BSPF. The shifting in the ionization energies is discussed in Refs. [37,38].

An analogous occupational probability is used for smoothing the truncation of the vibrational-rotational molecular partition function of deuterium molecules,

$$w_{n,J} = \exp\left[-\left(\frac{\Delta D_0}{-D_0 + (nh\nu + \varepsilon_J)}\right)^3\right] \tag{28}$$

where $\Delta D_0$ is the shifting in the dissociation energy which can be found from the configurational component of the free energy as

$$\Delta D_0 = -\left(\frac{\partial F_{hs}}{\partial N_{D_2}} - 2\frac{\partial F_{hs}}{\partial N_D}\right) \tag{29}$$

Coulomb's component of the configurational free energy and the excluding-volume correction to the bare nuclei free energy do not contribute to the lowering of dissociation energy as can be verified by replacing $F_{hs}$ by $F_C$ or $F_{bn}$ in the the right hand side of Eq. (29).

## VI– COMPUTATIONAL SCHEME

Letting $f$ be the fraction dissociated from the original number of molecules $N_0$, then the number of molecules at any time is $N_{D2}=(1-f)N_0$ and the number of neutral deuterium atoms is $N_D=2fN_0\alpha_0$ where $\alpha_0$ is the proportion of the neutral atoms, calculated from the ionization equilibrium, relative to the total number of atoms plus ions i.e., $\alpha_0=N_D/(N_D+N_{D+})$. Accordingly, the minimization equation of the free energy in the dissociation/recombination reaction (Eq. (3)) can be written in terms of $f$ as

$$\frac{\alpha_0^2 f^2}{(1-f)} = \frac{(Q_{tot,D}/V)^2}{4n_0\,(\overline{Q}_{tot,D_2}/V)} \times \exp\left[\frac{-1}{K_B T}\left(D_0 + \frac{\partial F_{Corr}}{\partial N_{D_2}} - 2\frac{\partial F_{Corr}}{\partial N_D}\right)\right]$$
$$\times \exp\left[+\frac{N_{D_2}}{Q_{D_2}}\left(\frac{\partial Q_{D_2}}{\partial N_{D_2}} - 2\frac{\partial Q_{D_2}}{\partial N_D}\right) + \frac{N_D}{Q_D}\left(\frac{\partial Q_D}{\partial N_{D_2}} - 2\frac{\partial Q_D}{\partial N_{D_2}}\right)\right] \quad (30)$$

where $n_0=N_0/V$ is the original number density of molecules. The partition functions in Eq. (30) are the conventional partition functions calculated using the respective ground states as references for the corresponding energies. The solution of the ionization equilibrium problem has been thoroughly discussed in other places [45-47].

For a constant temperature, $T$, and a specified mass density, $\rho$, the computational scheme used to solve the problem can be briefed to:

1- Start and read the temperature $T$, density $\rho$, and spectroscopic data

2- Find the number density of the originally non-dissociated gas molecules, $n_0 = \rho N_{Av}/M_{wt}$ where $M_{wt}$ is the molecular weight of the non-dissociated gas molecules and $N_{Av}$ is the Avogadro's number

3- Initialize $\alpha_0=1$ and set $\alpha_{0,old}=\alpha_0$

4- ***Iteratively,*** solve the chemical equilibrium problem (Eq. (30)) to find $f$

5- Set $n_{D,D+}$ (the number density of neutral atoms plus deuterons)=$2fn_0$

6- ***Iteratively,*** solve the ionization equilibrium problem to find each of $\alpha_0$ and $\alpha_{D+}$ (see Refs. [45-47])

7- Check convergence, ***if*** $|(\alpha_0-\alpha_{0,old})/\alpha_{0,old}|$ < predetermined tolerance (*epsilon)*, continue to calculate thermodynamic properties and write the results; ***else***

8- Set $\alpha_{0,old}=\alpha_0$ and go to 4. Repeat until convergence

In the above *simplified* scheme, details about iterative solution of the chemical equilibrium problem and the ionization equilibrium problem are disregarded here to simplify the readability of the scheme. For completeness, we provide a simplified schematic flowchart in Figure 1 below.

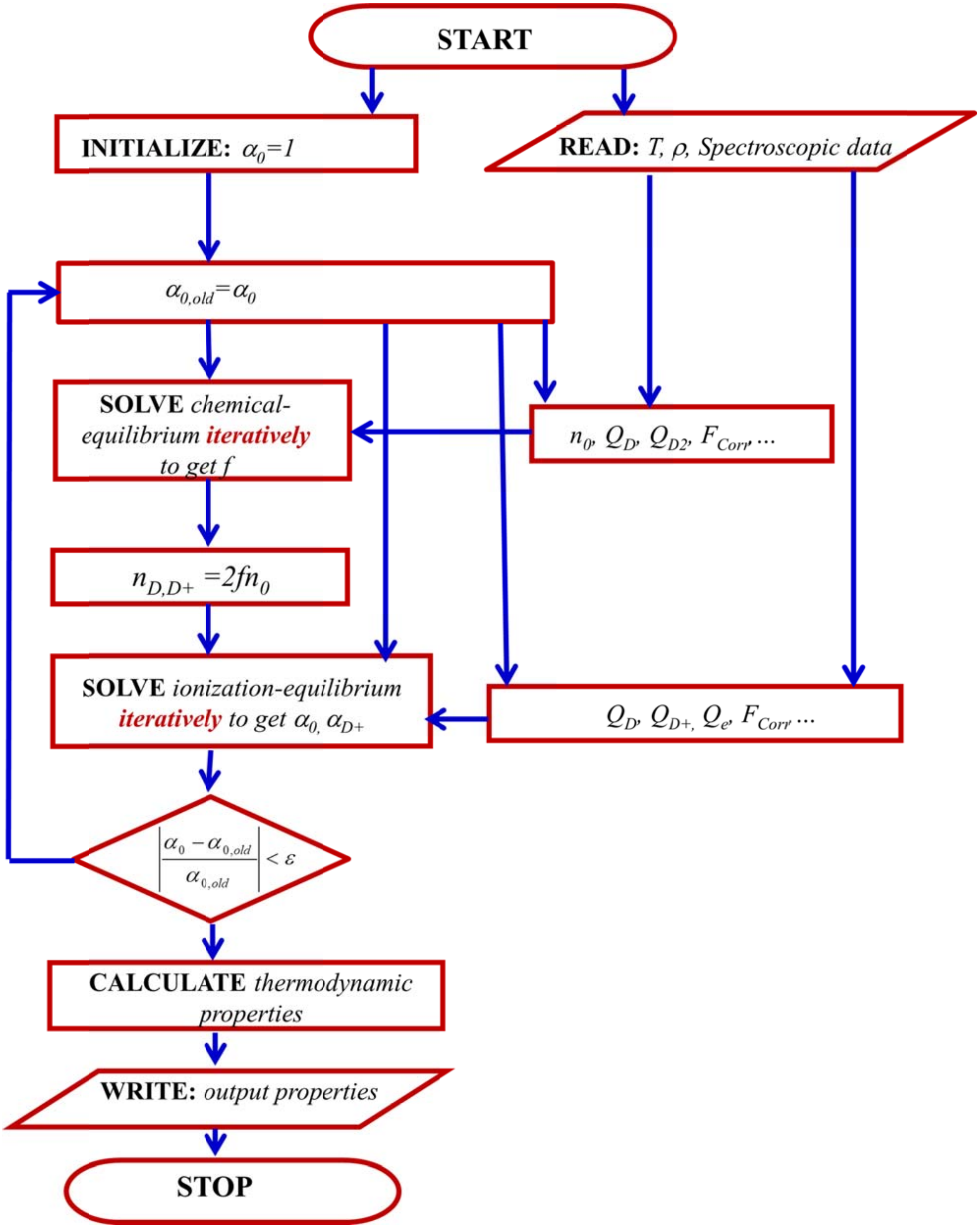

Figure 1. Simplified *Flowchart* of the used algorithm

## VII- RESULTS AND DISCUSSIONS

Isotherms of the degree of ionization, defined as ($n_{D+}/2n_0$,) where $n_0$ is the original number of molecules are presented in figure 2, for a wide range of densities (0.0001 g/cm$^3$ to 15 g/cm$^3$) and temperatures (2000 K  to 10$^6$ K). Isotherms are equally spaced on the logarithmic scale with $\Delta ln\, T = 0.31073$. Isotherms of higher temperatures are shown later where the free energy of radiation is taken into account. Some interesting features can be recognized from the figure:

a- Enhancement in ionization at intermediate densities (0.1 g/cm3 to10 g/cm$^3$) and relatively low temperatures (<45,000 K) due to Coulomb interactions. This enhancement in ionization is supressed at higher densities due to degeneracy of free electrons

b- Pressure enhanced ionization at extremely high densities at all temperatures due to hard core repulsion of extended species where equilibrium occupation numbers are shifted towards smaller volume particles (deuterons). If this effect was not considered, complete recombination between deuterons and electrons would be the result at extremely high density leading to neutral deuterium

c- The current predictions are exceptional in showing the correct physical behavior of non-intersecting isotherms in contrast to the results of many studies in the literature

d- Another distinct feature in the present results for the isotherms of the degree of ionization is that, for the shown range of densities, increasing the density at constant temperature enhances ionization though it does not lead to a fully stripped or completely ionized gas in contrast to most of the results in the literature which show a fully stripped nuclei at low temperaures and high densities. We believe that our results show the correct physical behavior, particularly when including short range hard core repulsion or equivalently excluding the volume of extended particles from the accessible volume for electrons and bare nuclei. Simply, if the results in the literature showing complete ionization of cold plasma at high density were correct, one would be left with a difficult to imagine physical situation where injecting material into the system (increasing density) leads to a decrease in the packing fraction, i.e. evacuation of the system. Some computational results in the literature (Graboske et al [42] for hydrogen and Kahlbaum and Forster [52] for helium) have shown such a decrease in the packing fraction with the increase of density at constant temperature due the shifting towards point-like electrons and bare nuclei which we believe are physically inappropriate. Our calculations of the packing fraction are shown in Fig. 3 where as can be seen from the figures isotherms of the packing fraction do not show negative slope though the slope gradually decrease towards zero, which is physically sensible.

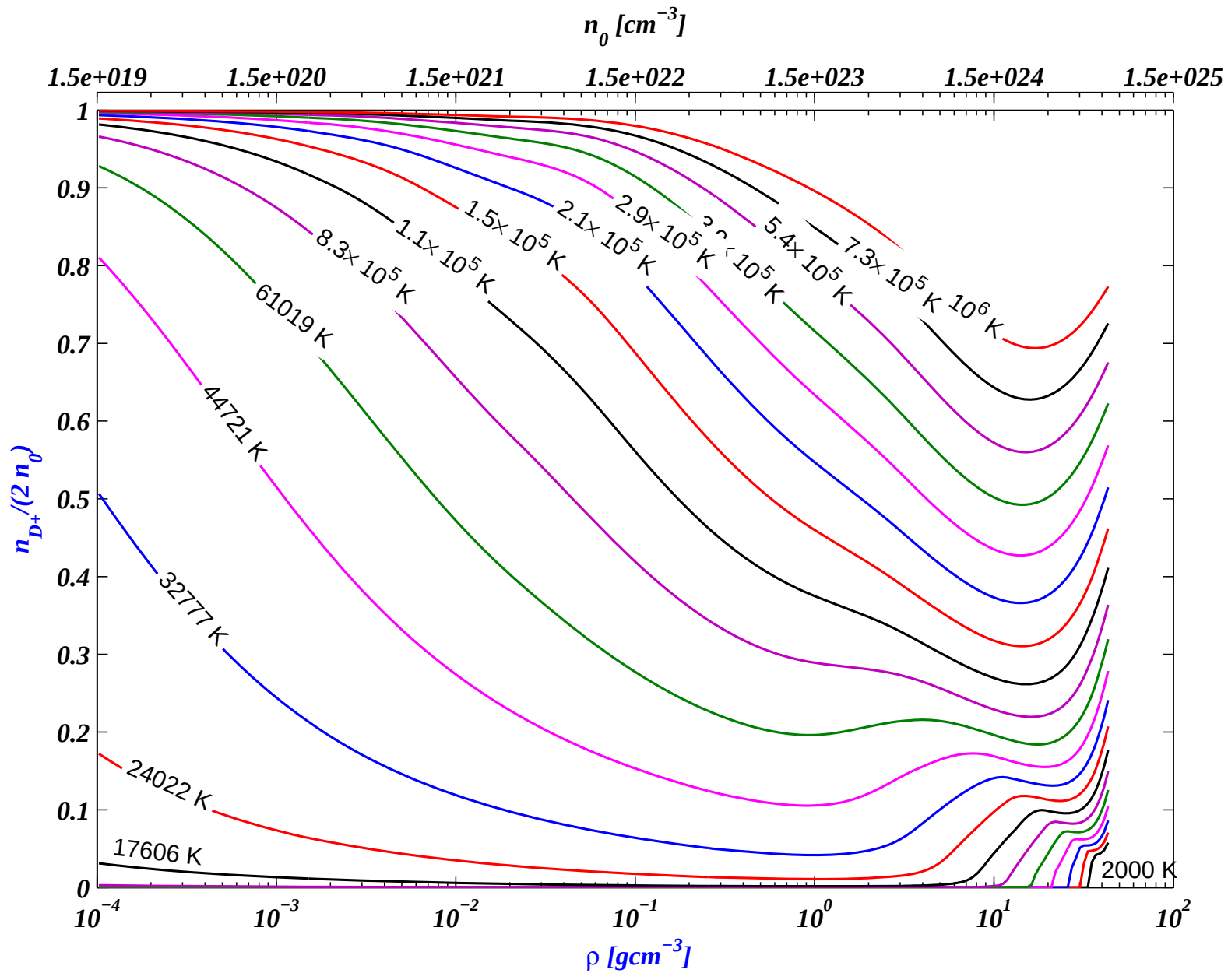

Figure 2. Degree of ionization, $n_{D+}/2n_0$, as a function of density for temperatures from 2,000 K to 1,000,000 K. Isotherms are equally spaced on the logarithmic scale with $\Delta ln\, T = 0.31073$.

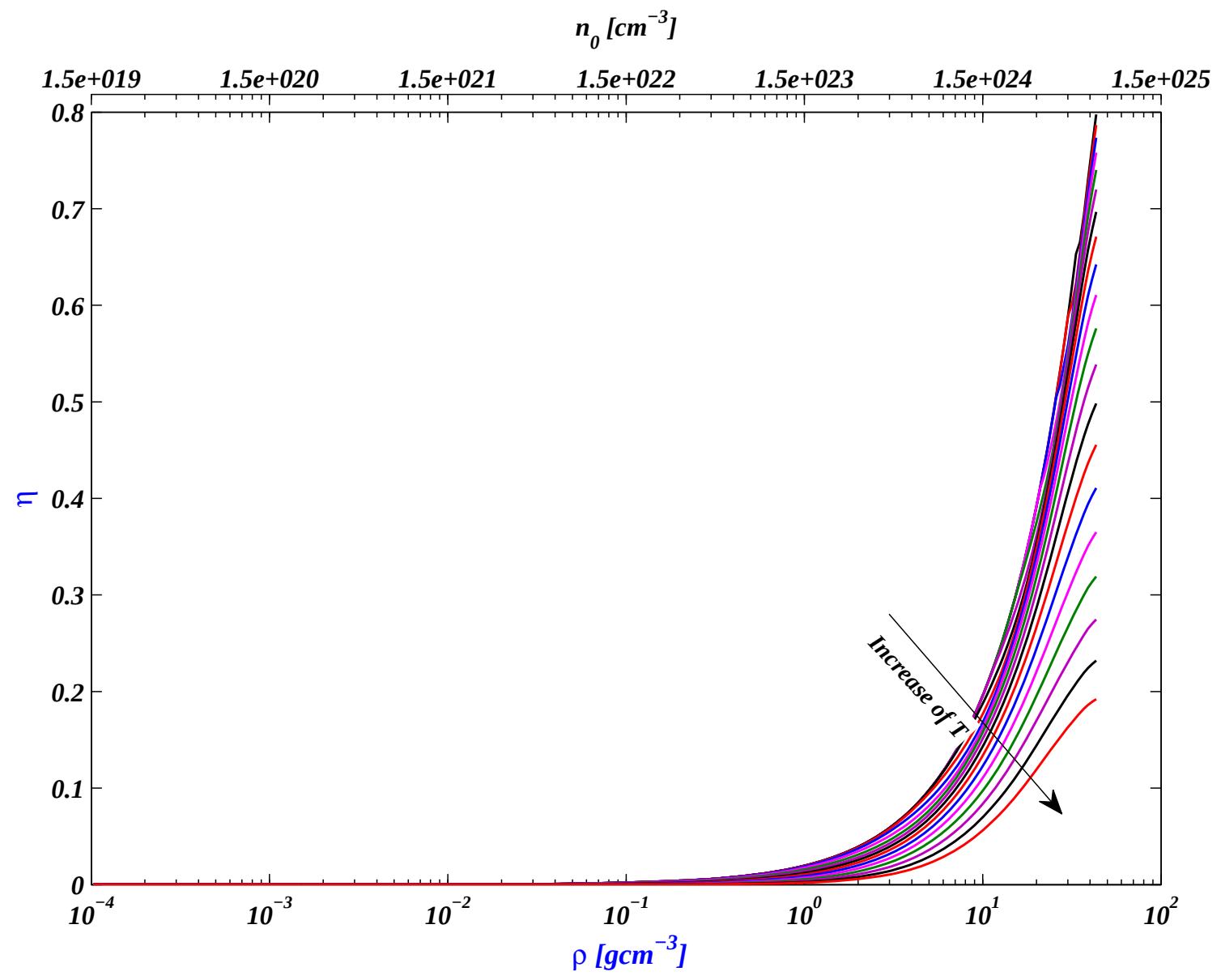

Figure 3. The packaging parameter, η, as a function of density for the same set of isotherms shown in Fig. 2.

Figure 4 shows the dissociation fraction, $f=(n_D+n_{D+})/2n_0$, for the same isotherms as Fig. 2. The common practice of increasing the dissociation with temperature (temperature dissociation) is obvious in the Figure from the sequence of isotherms. As the density increases from $\rho\approx10^{-4}$ to 1 g/cm$^3$ the probability of the formation of $D_2$ molecules increases due to the collision recombination of deuterium atoms and the chemical equilibrium is shifted towards the reverse reaction $D_2 \leftarrow 2D$ with the decrease of the degree of dissociation. At higher densities, strong nonideal correlation effects come into play, removing internal levels (lowering the dissociation energy) favoring molecular dissociation and the reaction is shifted towards the forward reaction $D_2 \rightarrow 2D$ with a strong increase in the dissociation fraction leading to complete pressure dissociation. It has to be noted that although complete cold ionization is not achievable in our calculation because it leads to a decrease in the packing fraction with the increase in density, complete cold dissociation is possible because of the equality of the volume of a molecule and the sum of the volumes of its constituent neutral atoms. Pressure dissociation appears at higher densities for all temperatures <.70,000 K. For temperatures >70,000 K one may consider the deuterium gas to be fully dissociated. Again, the current results are exceptional in showing non-intersecting isotherms in contrast to results in the literature. One final distict feature of the present results is that pressure dissociation occurs gradually over a range of densities and depends on the temperature.

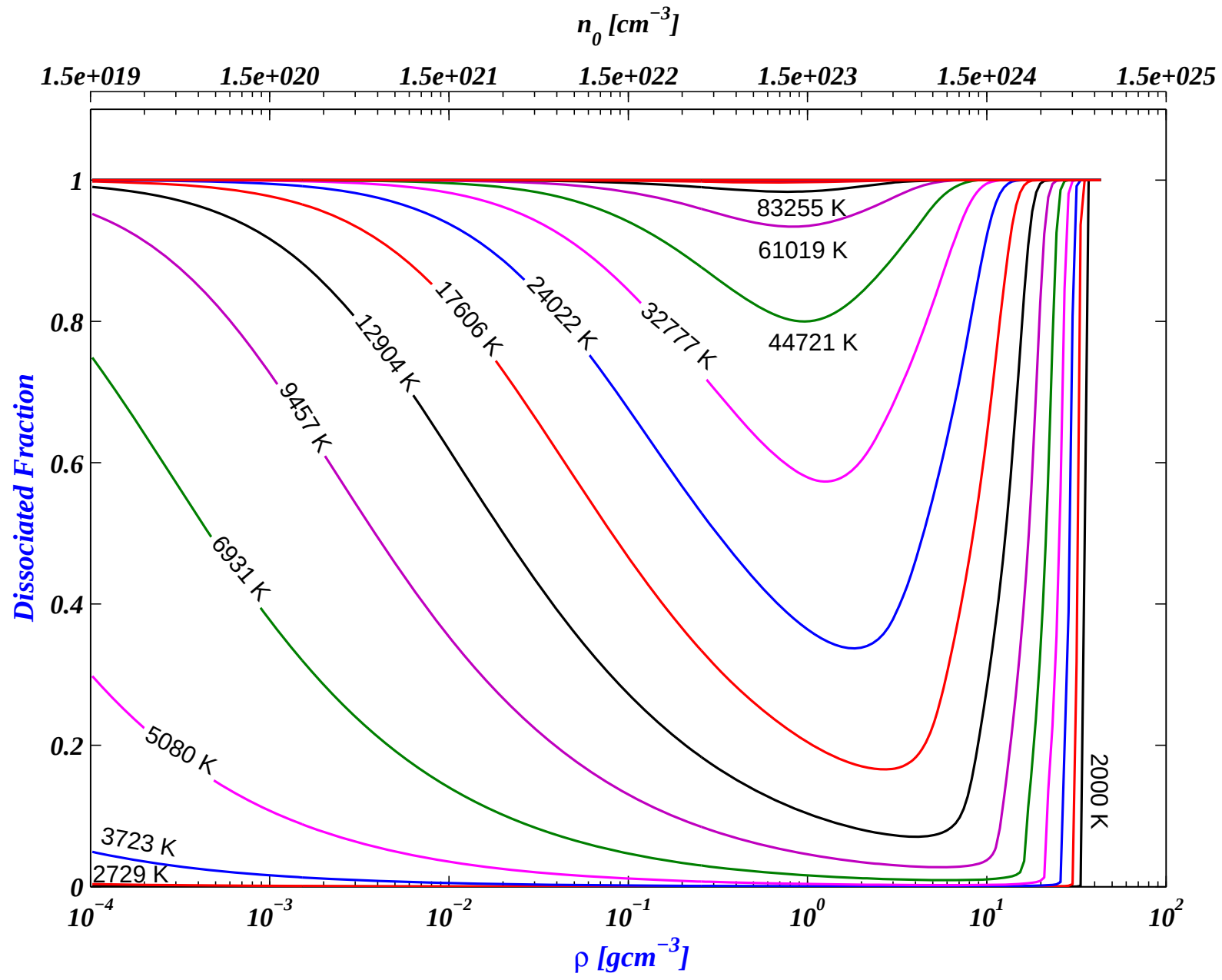

Figure 4. Dissociation Fraction, $f=(n_D+n_{D+})/2n_0$

Isotherms of the total pressure and of the specific internal energy of deuterium fluid are shown in figure 5 and figure 6, respectively. The pressure appears to behave ideally at low densities and/or high temperatures. However, deviations from this ideal behavior become clear at high densities and relatively low temperature. The negative values of the specific internal energy at low temperatures are indicative of the presence of a significant amount of dimers.

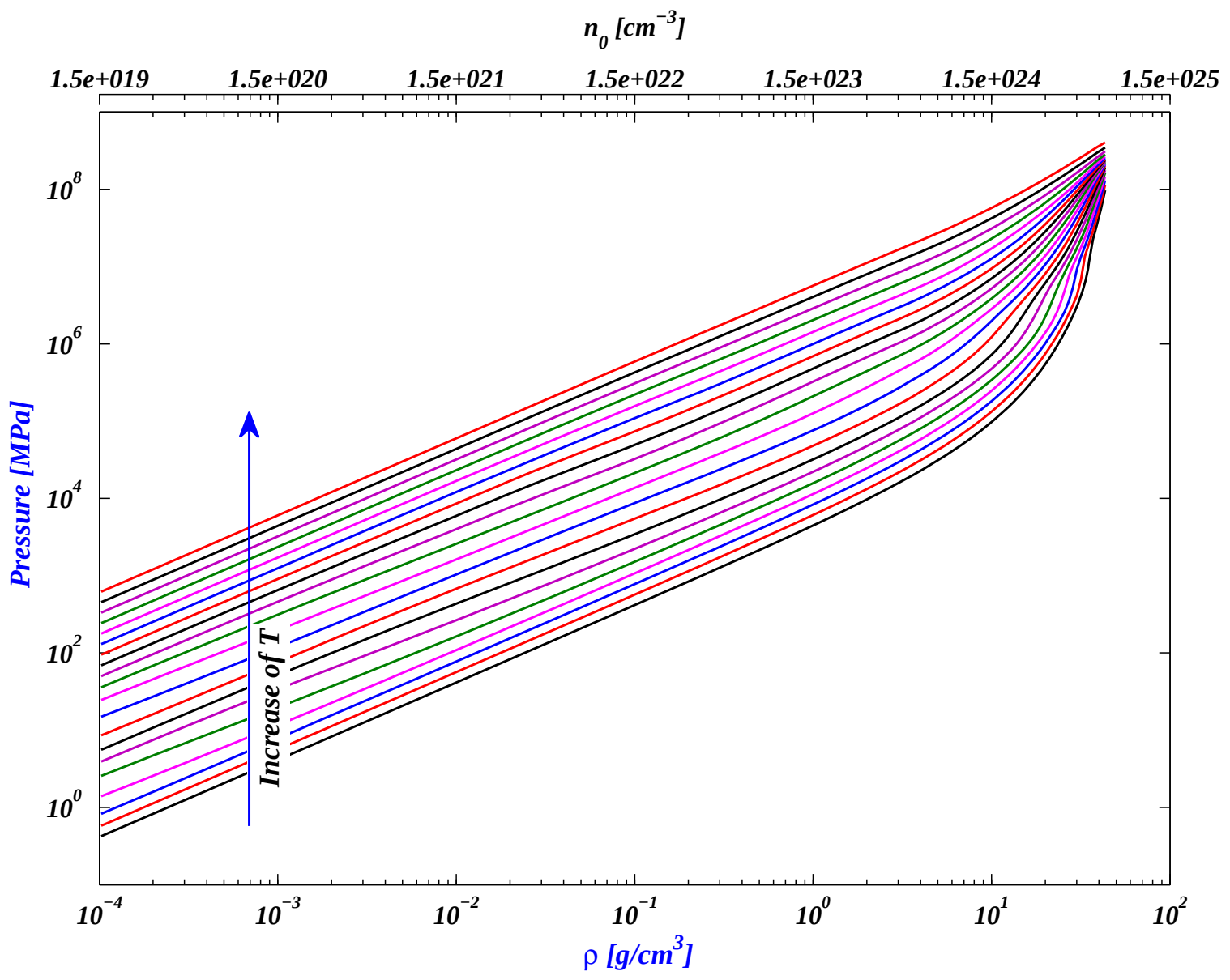

Figure 5. Pressure of fluid deuterium as a function of density for the same set of isotherms shown in Fig. 2.

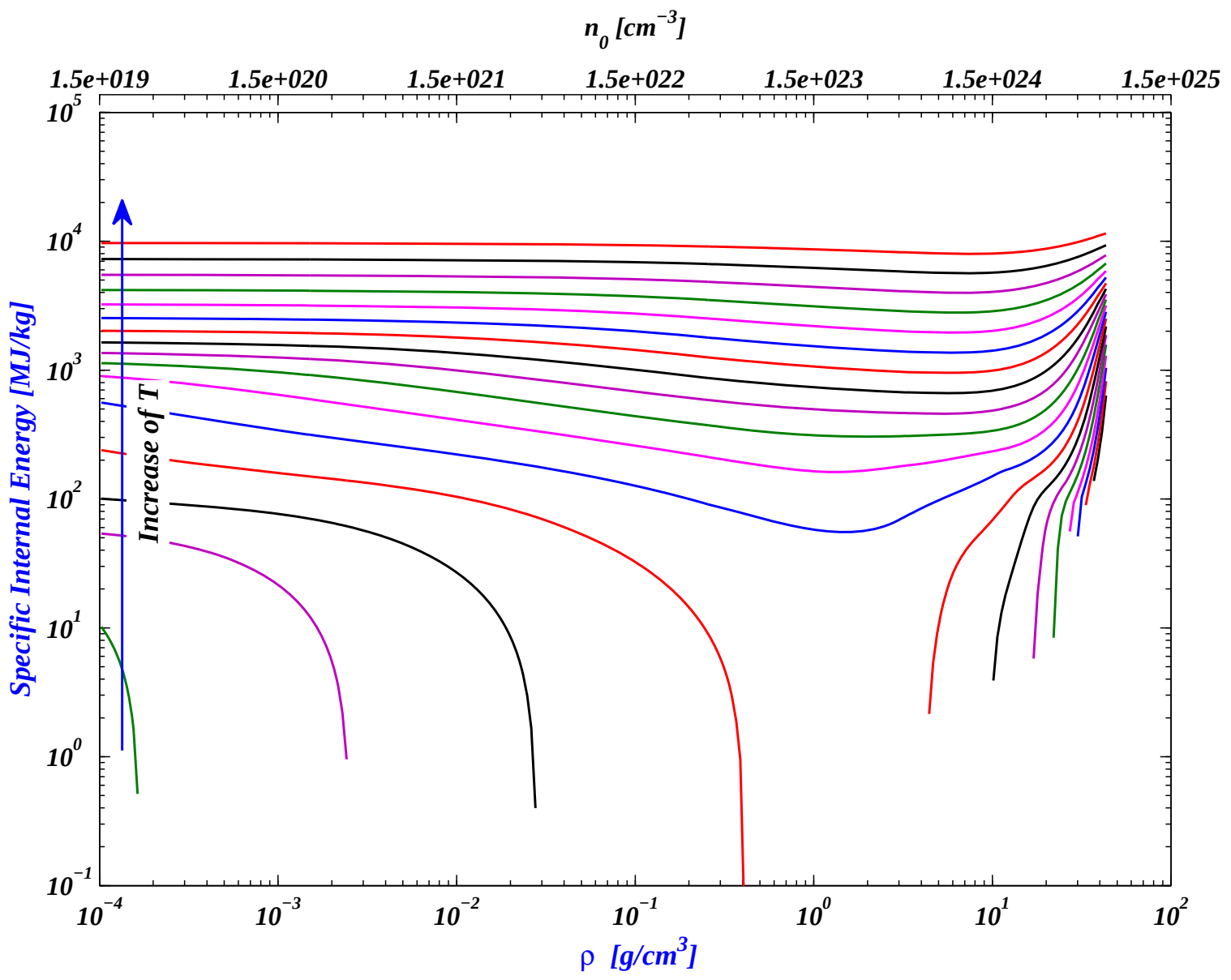

Figure 6. Isotherms of the specific internal energy of fluid deuterium (same set as in Fig. 2)

The effect of including the photon gas on the calculation of the pressure and specific internal energy of deuterium fluid is presented in figure 7 and figure 8, respectively. It is clear from the figures that while one can neglect the contribution of the photon gas to pressure and internal energy for T<1,000,000 K, this contribution is extremely important and plays the major part at high temperatures.

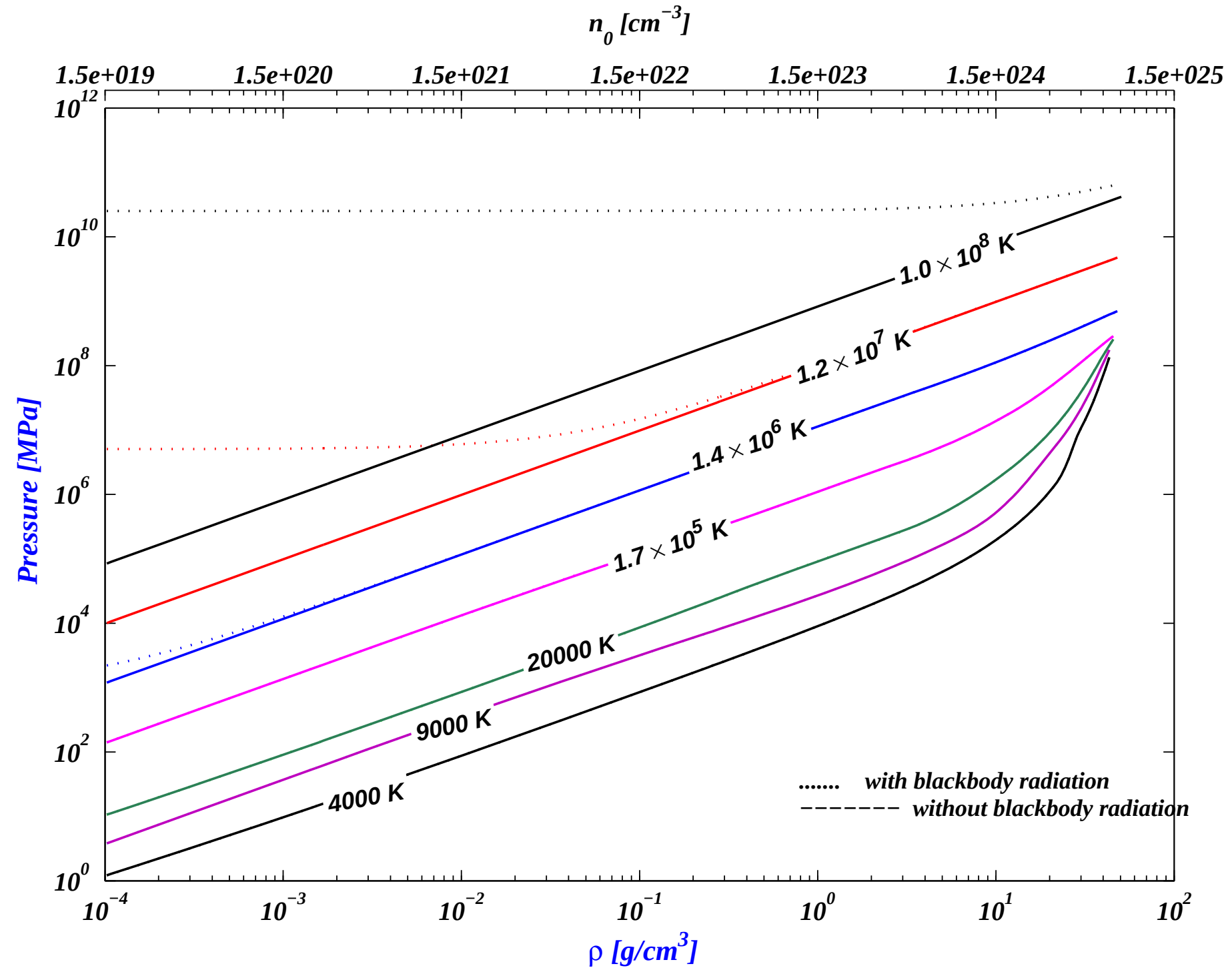

Figure 7. Effect of including radiation on pressure of fluid deuterium

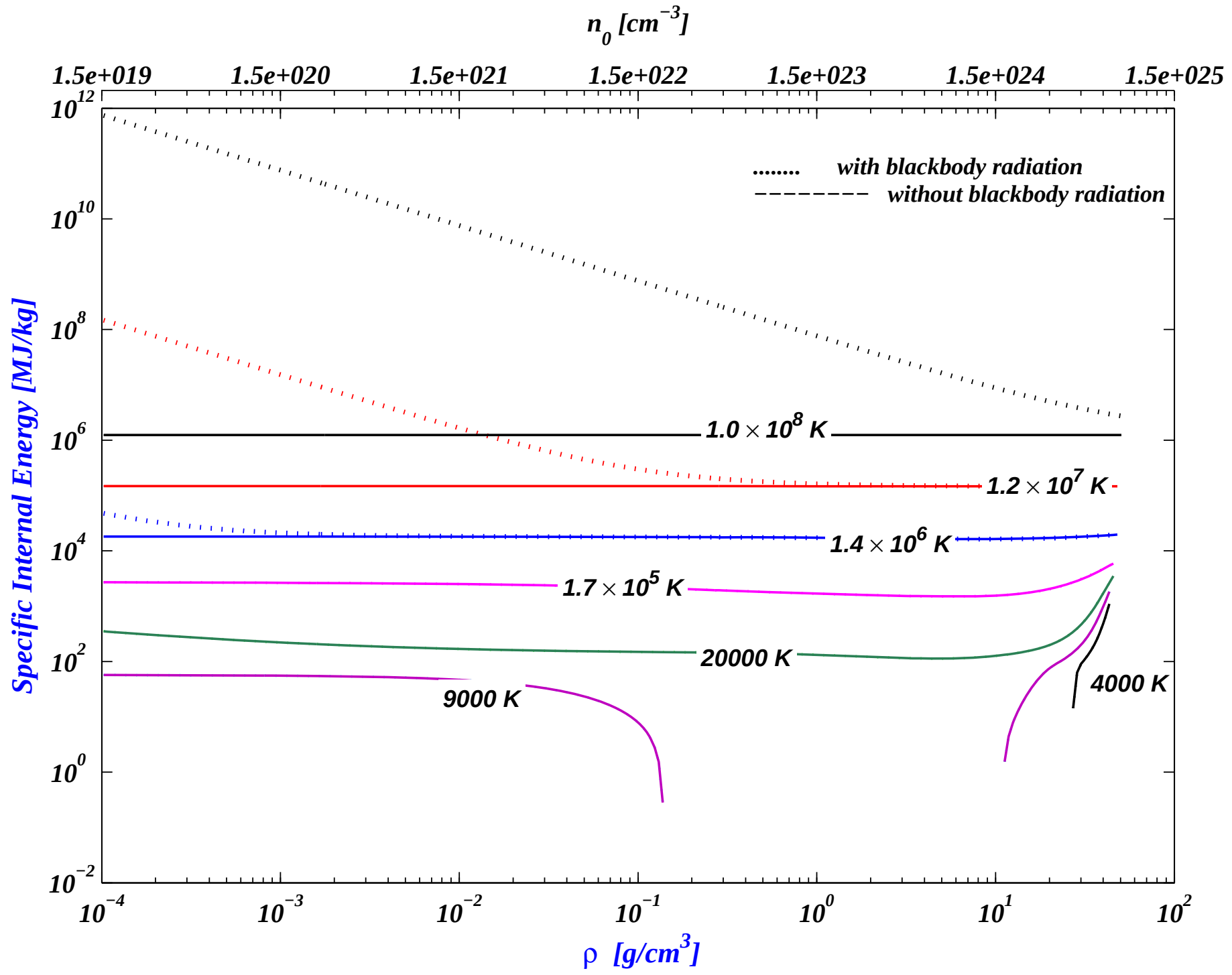

Figure 8. Effect of including radiation on internal energy of fluid deuterium

## VIII- CONCLUSIONS

A new model for strongly-coupled deuterium fluid is introduced and has been used to estimate the degree of ionization, dissociation fraction, equation-of-state and thermodynamic properties of deuterium over a wide range of temperatures and densities. An occupational probability is introduced and used to smoothly truncate the rotational-vibrational partition function which led to the pressure dissociation manifested in the computational results of the dissociation fractions.

Coulomb interactions and intensive hard-core repulsion (excluded-volume) strongly influence the equilibrium composition and thermodynamic properties.

Inclusion of the photon gas is shown to be necessary for temperatures >1,000,000 K. The present computational estimations do not suffer from the problem of intersecting isotherms found in other calculations in the literature and show a better and sensible physical behavior for the degree of ionization and packing fraction.